\documentclass[epj]{svjour}
\usepackage{graphics}
\begin{document}
\title{Pre-neutron-emission mass distributions for reaction $^{238}$U(n, f) up to 60 MeV}
\author{Xiaojun Sun\inst{1}\thanks{\emph{E-mail:} sxj0212@gxnu.edu.cn}
\and Chenggang Yu\inst{2} 
\and Ning Wang\inst{1}
\and Yongxu Yang\inst{1}
}                     
%
%
\institute{College of Physical Science and Technology, Guangxi Normal University, Guilin 541004, China \and Shanghai Institute of Applied Physics, Chinese Academy of Sciences, Shanghai 201800, China}
\date{Received: date / Revised version: date}
%
\abstract{
The pre-neutron-emission mass distributions for reaction $^{238}$U(n, f) up to 60 MeV are systematically studied with an empirical fission potential model. The energy dependence of the peaks and valleys of the pre-neutron-emission mass distributions is described by an exponential form based on the newly measured data. The energy dependence of evaporation neutrons before scission is also considered, which plays a crucial role for the reasonable description of the mass distributions. The measured data for the pre-neutron-emission mass distributions for reaction $^{238}$U(n, f) are reasonably well reproduced up to 60 MeV. The mass distributions at unmeasured energies are also predicted with this approach.
} 
\maketitle
\section{Introduction}
\label{intro}

Since the discovery of neutron-induced fission of Uranium in 1938 \cite{Hahn1939}, the neutron-induced fission is the subject of both theoretical and experimental studies. In the past, tremendous efforts have been focused on the low-energy actinide fission because of the particular importance for nuclear energy applications.
Nowadays, there is an increasing interest in studying neutron-induced fission of actinides at intermediate energies. It is motivated by nuclear data needs for new applications such as accelerator-driven system, thorium-based fuel cycle, and the next generation of exotic beam facilities. The pre-neutron-emission mass distribution is one of the most important quantities for neutron-induced fission. Its precise description is of great importance for both understanding the fission mechanism and the practical application. In addition, $^{238}$U is one of the most important actinides, and its disposal in spent fuel ($^{238}$U is up to 95\%) is an important feature of the utilization of nuclear power.

Although one can qualitatively describe the nuclear fission process as a deformation of a single nucleus, the exactly understanding the fission process or quantitatively predicting the pre-neutron-emission fragment mass distributions or product yields are still very elusive for the existing theories and models \cite{Randrup2011}. An international working group has studied the overall problem and recommended the assembly of the required nuclear data (including fission products) at intermediate incident neutron energies up to 150 MeV \cite{Lammer200801}. Compared with low-energy fission, the modeling of neutron-induced fission at intermediate energies is severely complicated by the fact that fission follows pre-equilibrium particle emission and competes with neutron evaporation \cite{Ryzhov2011}.

Several important theories and models \cite{Randrup2011,Goutte2005,Vanin1999,Karpov2001,Hu1999,Dubray2008,Younes2009,Moller2001,Moller2004,Moller2009,Moller2011,Brosa1990,Duijvestijn2001,Koning2007} have been developed for
understanding the fission mechanism or quantitatively calculating
the fragment mass distributions or fission product yields. These models are mainly focused on the dynamical processes. The systematic approaches
which consist of three to seven Gaussian functions have been developed for
quantitatively predicting the fragment mass distributions or
product yields \cite{Liu2008,Katakura2008,Wahl2008,Lammer2008}.

A combination method based on the driving potential from the Skyrme energy-density functional \cite{Liu2006,Wang2009} and the phenomenological fission potential is proposed in our previous work \cite{Sun2012}, and the experimental pre-neutron-emission mass distributions of neutron-induced actinide fission at low energies have been reasonably well reproduced. The present study is an extension of this combination method for reaction $^{238}$U(n, f) at incident energies up to 60 MeV.
This paper is organized as follows. In section 1, the combination method and the potential parameters are introduced in detail. In section 2, the comparisons of the calculated results and the measured data for reaction $^{238}$U(n, f) are presented and analyzed. A simple summary is also given in the this section.

\section{The combination method and the potential parameters}
\label{sec:1}

\begin{figure}
\resizebox{0.5\textwidth}{!}{%
\includegraphics{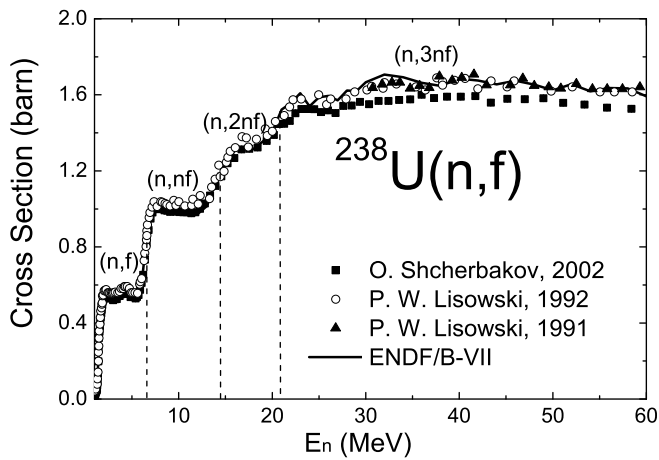}
}
\caption{Fission cross section of reaction $^{238}$U(n, f)
for incident neutron energies from threshold energy to 60 MeV. The experimental data are obtained from Refs. \cite{Shcherbakov2002}(squares), \cite{Lisowski1992}(circles) and \cite{Lisowski1991}(triangles), respectively. The solid line denotes the evaluated results of ENDF/B-VII, and the dash lines label the incident energy regions corresponding to the different multi-chance fission channels such as (n, f), (n, nf), (n, 2nf) and (n, 3nf), respectively. }
\label{CS}
\end{figure}

The sequential products of neutron-induced binary fission are elaborated on Refs. \cite{Sun2012,Madland2006}.
A combination method for calculating the pre-neutron-emission mass
distributions of neutron-induced actinide fission at low energies
has been proposed in our previous work \cite{Sun2012}. In this model, the pre-neutron-emission mass distributions are described as
\begin{equation}\label{eq1}
P(A)=C\exp[-U(A)].
\end{equation}
Where $C$ is the normalization constant, and the variable $A$
denotes the mass number of the primary fragment. The
phenomenological fission potential $U(A)$ is described by three harmonic-oscillator functions, i.e.,
\begin{equation}\label{eq2}
    U(A)= \left\{\begin{array}{l l}
        \displaystyle u_1(A-A_1)^2               & A\leq a      \\
        \displaystyle -u_0(A-A_0)^2+R            & a\leq A\leq b \\
        \displaystyle u_2(A-A_2)^2               & A\geq b.      \\
    \end{array}\right.
\end{equation}
Where, $A_1$ and $A_2$ are the positions of the
light and heavy fragment peaks of the pre-neutron-emission mass
distributions, respectively. $A_0$ denotes the corresponding
position for symmetric fission. The fission potential parameters $u_1, u_0, u_2, a, b$ and $R$, which are the functions of $A_0, A_1$ and $A_2$, have been uniquely derived as Eq. (6) - Eq. (9) in our previous paper \cite{Sun2012}.

\begin{figure}
\resizebox{0.5\textwidth}{!}{%
\includegraphics{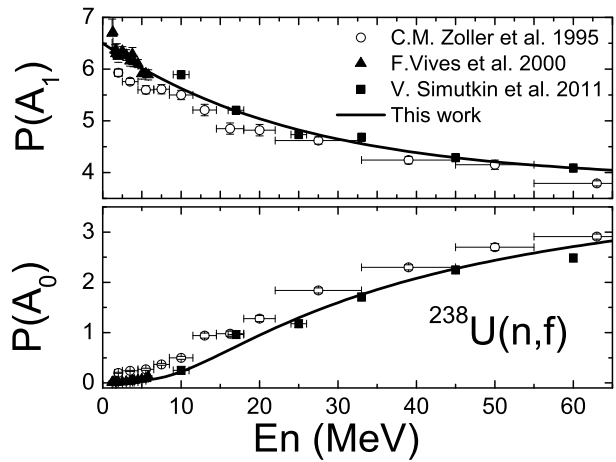}
}
\caption{Peak $P(A_1)$ and valley $P(A_0)$ of
the pre-neutron-emission mass distributions for reaction $^{238}$U(n, f) as a function of incident neutron energy. The experimental data are derived from the white neutron beam (circles) \cite{Zoller1995}, monoenergetic neutron (triangles) \cite{Vives2000} and the quasi-monoenergetic neutron (squares) \cite{Ryzhov2011,Simutkin2011}. The solid lines denote the results of this work. }\label{PA}
\end{figure}
\begin{table*}
\caption{The positions ($A_1, A_2)$ for the mass number of the light and heavy fragments mass distributions for reaction $^{238}$U(n, f) at different incident energy regions.}
\label{table1}
\begin{tabular}{llllllll}
\hline\noalign{\smallskip}
$E_n$ (MeV) &9-11       &16-18      &24-26      &33         &45         &60         &Ref.\\
\noalign{\smallskip}\hline\noalign{\smallskip}
Experiment  &(99, 138)  &(99, 138)  &(98, 138)  &(99, 137)  &(99, 137)  &(99, 136)  &\cite{Simutkin2011} \\
TALYS       &(99, 139)  &(99, 138)  &(99, 138)  &(98, 137)  &(98, 137)  &(98, 136)  &\cite{Koning2007}\\
This work   &(99, 139)  &(99, 138)  &(99, 137)  &(99, 137)  &(99, 137)  &(99, 137)  &\\
\noalign{\smallskip}\hline
\end{tabular}
\end{table*}

\begin{figure*}
\resizebox{1\textwidth}{!}{%
\includegraphics{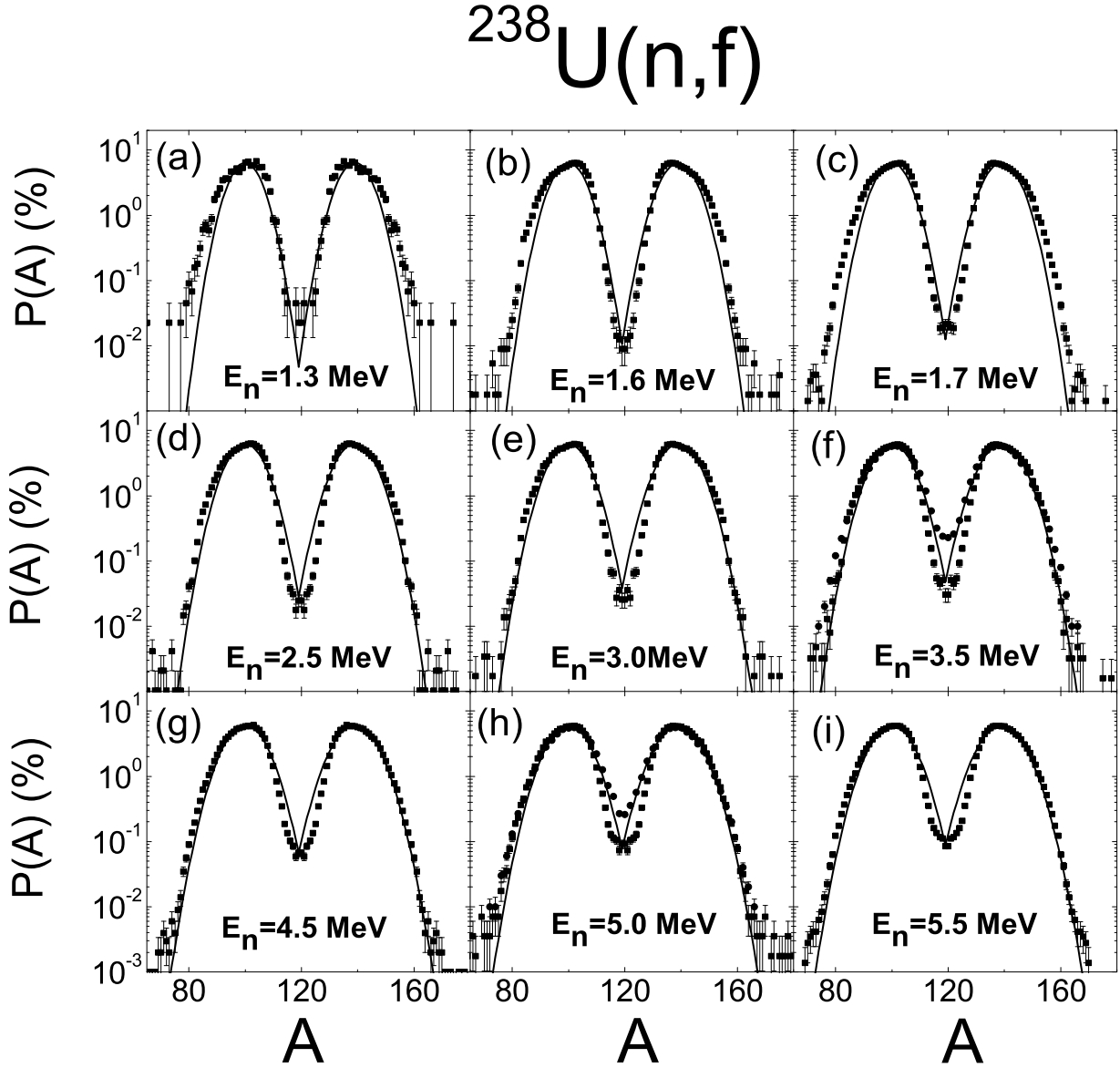}
}
\caption{Pre-neutron-emission mass distributions
at incident energies from 1.3 to 5.5 MeV for reaction $^{238}$U(n, f). The
scattered symbols denote the experimental data, which are taken from Ref. \cite{Vives2000} (squares, measured by the monoenergetic neutron)
and from Ref. \cite{Zoller1995} (circles, measured by the white neutron beam), respectively.}\label{dist1}
\end{figure*}

A particular attention should be payed that these parameters are closely relative to the evaporation neutrons before scission at different incident energies. For reaction $^{238}$U(n, f) at low incident energies ($E_n \leq 6.5$ MeV), the
positions $A_1$ and $A_2$ are obtained from the nucleus-nucleus
driving potential of the fissile nucleus $^{239}$U \cite{Liu2006,Wang2009}.
With the incident neutron energy increasing, the excitation energy of the
compound nucleus will become higher, and a few neutrons will be
evaporated before scission. The numbers
of the evaporation neutron can be derived from the corresponding multi-chance fission cross sections. Therefore, the fission cross
sections of reaction $^{238}$U(n, f) have been investigated as shown in Fig. \ref{CS}. The scattering dots denote the experimental data derived from Refs. \cite{Shcherbakov2002,Lisowski1992,Lisowski1991}, and the solid line does the evaluated results of ENDF/B-VII, which is recommended as the standard cross sections. The dash lines denote the incident energy regions corresponding to the different multi-chance fission channels as labeled (n, f), (n, nf), (n, 2nf) and (n, 3nf), respectively. From Fig. \ref{CS}, one established that the number $\tilde{n}(E_n)$ of evaporation neutrons before scission can be
roughly expressed as follow
\begin{equation}\label{eq3}
 \tilde{n}(E_n)=\left\{\begin{array}{llllll}
        \displaystyle 0,          &  & E_{\rm th}\leq E_n \leq6.5 & \textrm{MeV}      \\
        \displaystyle 1,          &  & 6.5< E_n \leq 14.5 & \textrm{MeV}      \\
        \displaystyle 2,          &  & 14.5<E_n\leq21.5 & \textrm{MeV}      \\
        \displaystyle 3           &  & 21.5<E_n\leq60 & \textrm{MeV}.      \\
    \end{array}\right.
\end{equation}
Where, $E_{\rm th}$ is the threshold energy for
$^{238}$U(n, f) reaction. Eq. (\ref{eq3}) is consistent with the result at low incident energies as shown in Ref. \cite{Sun2012}.

\begin{figure*}
\resizebox{1\textwidth}{!}{%
\includegraphics{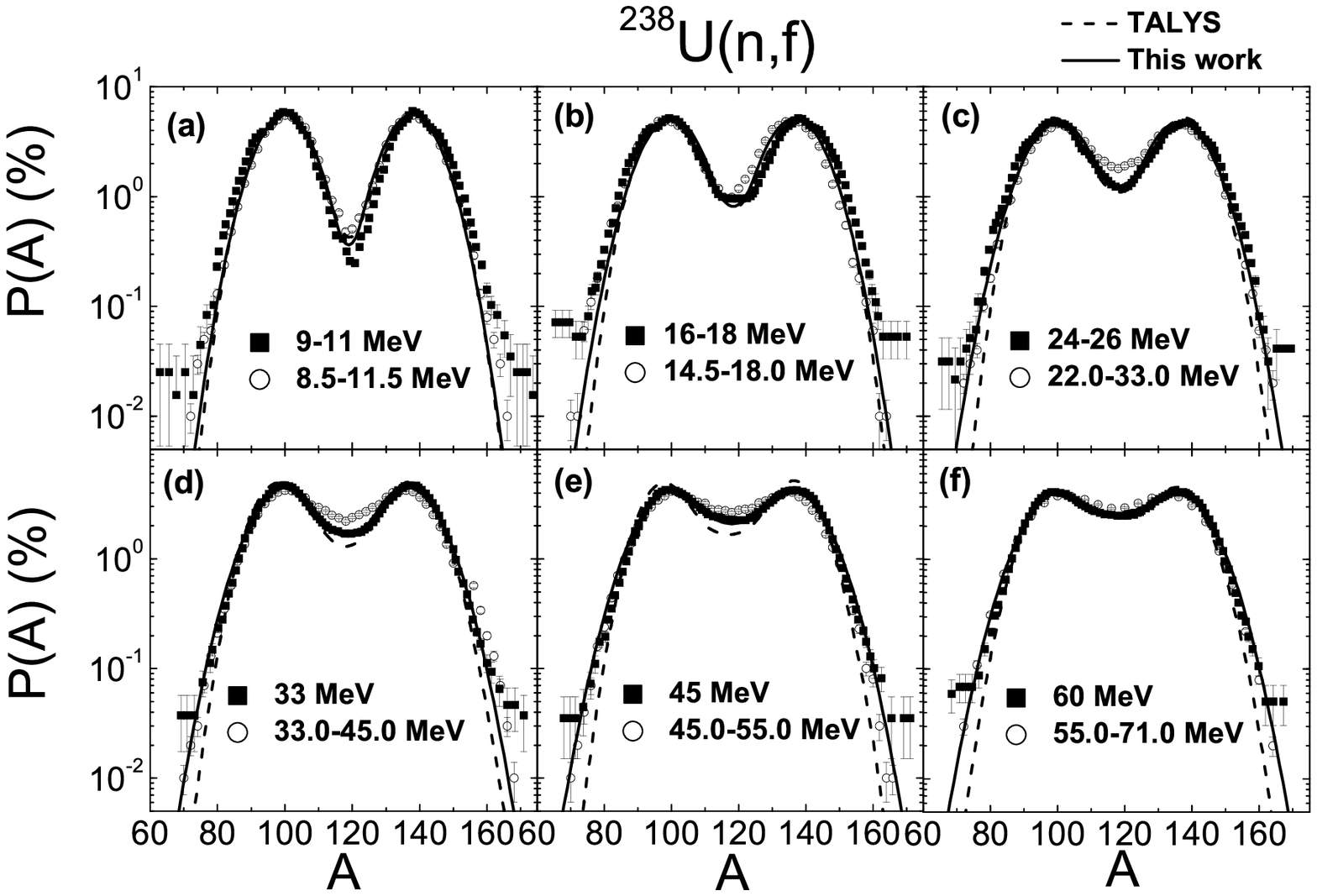}
}
\caption{Pre-neutron-emission mass distributions
at incident energies from 10 to 60 MeV for reaction $^{238}$U(n, f). The
scattered symbols denote the experimental data, which are taken from Ref. \cite{Ryzhov2011,Simutkin2011} (squares, measured by the quasi-monoenergetic neutron)
and from Ref. \cite{Zoller1995} (circles, measured by the white neutron beam), respectively. The dash and solid
curves denote the calculated results of TALYS code \cite{Koning2007} and in this work, respectively.}\label{dist2}
\end{figure*}

It is assumed that a compound nucleus $A_{\rm CN}$ after evaporating neutrons $\tilde{n}(E_n)$ separates into a pair of fragments in the fission process, so the mass number of the fissile nucleus is $A_{FN}=A_{CN}-\tilde{n}(E_n)$ at different incident energy regions. For reaction $^{238}$U(n, f), the fissile nuclei are $^{239}$U, $^{238}$U, $^{237}$U and $^{236}$U, respectively, at different incident energy regions as listed in Eq. (\ref{eq3}).
Based on the nucleus-nucleus potential with the Skyrme energy-density functional \cite{Liu2006,Wang2009}, the driving potentials of these fissile systems are studied considering the deformations of fragments. One sees that these driving potentials generally show a valley at
$A\sim$140 for the mass distributions of heavy fragments, as elaborated Fig. 1 in Ref. \cite{Sun2012}. It is should be noted that the driving potentials are only derived from the ground state or low excited energies of the fragments. However, the fissile nuclei still hold highly excited energies after evaporating neutrons at different incident energy regions.
So the position $A_2$ of the heavy fragment peaks, as well as $A_1$ of the light fragment peaks and $A_0$ of the symmetrical fission, should be modified as
\begin{equation}\label{eq4}
\left\{\begin{array}{llllll}
        \displaystyle A_2=A_2^{g.s.}-\tilde{n}(E_n),      \\
        \displaystyle A_1=A_{FN}-A_2,     \\
        \displaystyle A_0=A_{FN}/2.     \\
       \end{array}\right.
\end{equation}
Where $A_2^{g.s.}\simeq 140$ denotes the lowest position of the driving potential derived from the ground state of the fragments. These results of Eqs. (\ref{eq3}) and (\ref{eq4}) agree exactly with the positions of the maximal mass distributions of the heavy fragments measured by the quasi-monoenergetic neutrons beam from 10 MeV up 60 MeV \cite{Ryzhov2011,Simutkin2011} as listed in Table 1. For comparison, the results of the famous TALYS code \cite{Koning2007} are also listed in this Table.

Based on monoenergetic experimental data \cite{Vives2000} and the quasi-monoenergetic experimental data \cite{Ryzhov2011,Simutkin2011}, the heights $P(A_0)$ and
$P(A_1)$ of the valleys and peaks of the pre-neutron-emission mass
distributions have been fitted as the functions of incident
neutron energy. For reaction $^{238}$U(n, f) up to 60 MeV, the energy dependence of $P(A_1)$ and $P(A_0)$ is written as
\begin{eqnarray}\label{eq5}
P(A_1) &=& 3.850+2.600 e^{-0.040E_n},\nonumber\\
P(A_0) &=& 0.044+4.581 e^{-32.465/E_n}.
\end{eqnarray}
The results of the measurement, including from the white neutron beam \cite{Zoller1995}, monoenergetic neutron \cite{Vives2000} and the quasi-monoenergetic neutron \cite{Ryzhov2011,Simutkin2011}, and the calculation are shown in Fig. \ref{PA}. So the parameter $R$ in Eq. (2) can be derived easily through Eq. (\ref{eq6}) as shown
\begin{equation}\label{eq6}
R=\textmd{ln}\frac{P(A_1)}{P(A_0)}.
\end{equation}
Furthermore, Eq. (\ref{eq5}) approximatively equals the results of Ref. \cite{Sun2012} at low incident energy ($E_n \leq 6.5$ MeV). One can see that the values of $P(A_1)$ and $P(A_0)$ exponentially change with the incident energies in general, which could provide some useful information at unmeasured energies.

\begin{figure}
\resizebox{0.55\textwidth}{!}{%
\includegraphics{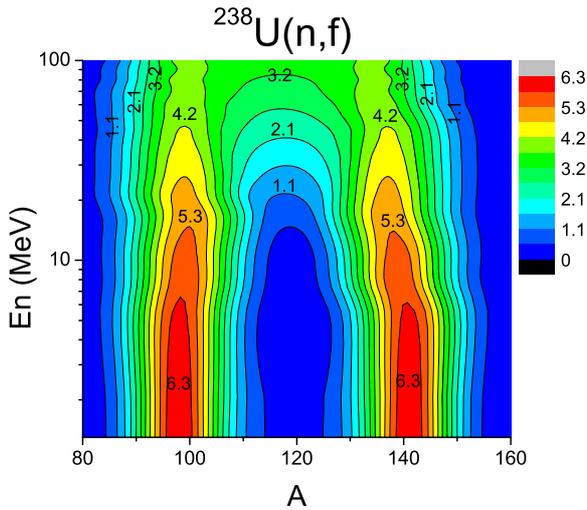}
}
\caption{(Color online) The calculated pre-neutron-emission mass distributions (\%)
at incident energies from threshold energy up to 100 MeV for reaction $^{238}$U(n, f) involved the fragment mass number $A$ and the incident energy $E_n$.}\label{dist3}
\end{figure}

\subsection{Results and analysis}
\label{sec:2}

In this work, the evaporation neutrons $\tilde{n}(E_n)$ at different incident energy regions are derived from the fission cross sections in multi-chance fission channels as shown in Fig. \ref{CS}. In terms of the evaporation neutrons $\tilde{n}(E_n)$, the positions $A_2$ of the heavy fragment peaks of the pre-neutron-emission mass distributions are determined. So the positions $A_1$ of the light fragment peaks can be also obtained easily. Combined the heights $P(A_0)$ and
$P(A_1)$ of the valleys and peaks of the pre-neutron-emission mass
distributions as shown in Fig. \ref{PA}, the parameter $R$ is also obtained in terms of Eq. (\ref{eq6}). So the pre-neutron-emission mass
distributions can be calculated using Eq. (\ref{eq1}) and (\ref{eq2}) up to 60 MeV, as shown in Fig. \ref{dist1} and \ref{dist2}. Fig. \ref{dist1} shows the pre-neutron-emission mass
distributions of reaction $^{238}$U(n, f) at low incident energies from 1.3 MeV to 5.5 MeV, and one can see that this results agree with the previous results of Ref. \cite{Sun2012}. Fig. \ref{dist2} shows the calculated results up to 60 MeV. In this figure, the scattered symbols denote the experimental data, which are taken from Ref. \cite{Ryzhov2011,Simutkin2011} (squares, measured by the quasi-monoenergetic neutron)
and from Ref. \cite{Zoller1995} (circles, measured by the white neutron beam), respectively. The solid curves
denote the calculated results in this work. The dash curves denote the results calculated by TALYS code \cite{Koning2007}.
One can see that the experimental data of reaction $^{238}$U(n, f) can be reproduced well at different incident neutron
energies from threshold energy up to 60 MeV. It indicates that the method combined the driving potential with phenomenological fission potential is reasonable to describe the pre-neutron-emission mass
distributions of reaction $^{238}$U(n, f) up to 60 MeV.

Fig. 5 gives the contours of the predicted pre-neutron-emission mass distributions of reaction $^{238}$U(n, f) from threshold energy to 100 MeV. One can see that several distinct characters of the pre-neutron-emission mass distributions can be reasonably reproduced: 1) the double bump shape; 2) the increase of the valley heights, as well as the decrease of the peak heights, with the incident energy increasing; 3) the position $A_2$ of the heavy fragment peak locates roughly 140 at low energies, and gradually decreases because of the evaporation neutrons before scission at $E_n>$6.5 MeV. Contrarily, the position $A_1$ of the light fragment peak always locates roughly 99 from threshold energy up to 100 MeV. This implies that the combination method in this paper can provide some additionally useful information for the intermediate energies neutron induced actinides fission.
\\

\textbf{Acknowledgements}

We thank our colleagues L. Ou and M. Liu for some valuable suggestions. This work was
supported by Guangxi University Science and Technology Research Projects (Grant No. 2013ZD007), Guangxi Natural Science Foundation (Grant No. 2012GXNSFAA053008),  the Th-based Molten Salt Reactor Power System of the Strategic Pioneer Science and Technology Projects from the Chinese
Academy of Sciences, and National Natural Science Foundation of China (Grants No. 11265004).

%
%

\end{document}